\documentclass[aip, jmp,amsmath,amssymb,preprint,]{revtex4-1}
\DeclareMathAlphabet{\mathpzc}{OT1}{pzc}{m}{it}
\usepackage{graphicx}% Include figure filesg
\usepackage{amsfonts}
\usepackage[normalem]{ulem}
\usepackage{color}
\usepackage{bm}
\usepackage{comment}
\usepackage{ulem}
\def\ie{{\it i.e.,\ }}
\def\etal{{\it et al.\ }}

\def\up#1{\raise1mm\hbox{$\!\!^{#1}$}} 
\def\upp#1{\raise2mm\hbox{$\!\!\!\!^{#1}$}} 
\newcommand{\Deqn}[1]{{Eq.~(\ref{#1})}}

\newcommand{\beq}{\begin{equation}}
\newcommand{\be}{\begin{equation}}
\newcommand{\eeq}{\end{equation}}
\newcommand{\ee}{\end{equation}}
\newcommand{\bea}{\begin{eqnarray}}
\newcommand{\eea}{\end{eqnarray}}
\newcommand{\bal}{\begin{align}}
\newcommand{\eal}{{\end{align}}}

\newcommand{\beaa}{\begin{eqnarray*}} 
\newcommand{\eeaa}{\end{eqnarray*}}

\newcommand{\bsube}{\begin{subequations}}
\newcommand{\esube}{\end{subequations}}

\newcommand{\eq}{\ \ \ \,}

\newcommand{\cost}{\cos\theta}
\newcommand{\sint}{\sin\theta}

\newcommand{\sSlmg}{\,_sS_{\ell,m}^\gamma}
\newcommand{\sTlmg}{\,_sT_{\ell,m}^\gamma}

\newcommand{\sYlm}{\,_sY_{\ell,m}}
\newcommand{\sYlpm}{\,_sY_{\ell+1,m}}
\newcommand{\sYlmm}{\,_sY_{\ell-1,m}}
\newcommand{\spYlm}{\,_{s+1}Y_{\ell,m}}
\newcommand{\spYlpm}{\,_{s+1}Y_{\ell+1,m}}
\newcommand{\spYlmm}{\,_{s+1}Y_{\ell-1,m}}

\newcommand{\sXlm}{\,_sX_{\ell,m}}
\newcommand{\sXlpm}{\,_sX_{\ell+1,m}}
\newcommand{\sXlmm}{\,_sX_{\ell-1,m}}
\newcommand{\spXlm}{\,_{s+1}X_{\ell,m}}
\newcommand{\spXlpm}{\,_{s+1}X_{\ell+1,m}}
\newcommand{\spXlmm}{\,_{s+1}X_{\ell-1,m}}

\newcommand{\sbllpm}{\,_{s}b_{\ell,\ell+1}^m}
\newcommand{\sbllmm}{\,_{s}b_{\ell,\ell-1}^m}

\newcommand{\g}{\gamma}

\newcommand{\sSlmgq}{\,\;S_{\ell,m\!\!\!\!\!\!\!\!\!\!\!\!\!\!\!\!\!\!s\,\,\,\,\,\,\,\,\,\,\,\,\,\,\,}^\gamma}
\newcommand{\sYlmq}{\,\;Y_{\ell,m\!\!\!\!\!\!\!\!\!\!\!\!\!\!\!\!\!s\,\,\,\,\,\,\,\,\,\,\,\,\,\,}^{{\color{white} \gamma}}}
\newcommand{\ssquarelmgq}{\,\;\hat{\square}_{\ell,m\!\!\!\!\!\!\!\!\!\!\!\!\!\!\!\!\!\!\!s\,\,\,\,\,\,\,\,\,\,\,\,\,\,\,}^\gamma}
\newcommand{\sElmgq}{\,\;E_{\ell,m\!\!\!\!\!\!\!\!\!\!\!\!\!\!\!\!\!\!\!s\,\,\,\,\,\,\,\,\,\,\,\,\,\,\,}^\gamma}
\newcommand{\spmSlmgq}{\,\;S_{\ell,m\!\!\!\!\!\!\!\!\!\!\!\!\!\!\!\!\!\!\!\!\!\!\!\!\!s\pm1\,\,\,\,\,\,\,\,\,\,\,\,\,\,\,}^\gamma}
\newcommand{\spmsquarelmgq}{\,\;\hat{\square}_{\ell,m\!\!\!\!\!\!\!\!\!\!\!\!\!\!\!\!\!\!\!\!\!\!\!\!\!\!s\pm1\,\,\,\,\,\,\,\,\,\,\,\,\,\,\,\,}^\gamma}
\newcommand{\sTlmgq}{\,\;T_{\ell,m\!\!\!\!\!\!\!\!\!\!\!\!\!\!\!\!\!\!s\,\,\,\,\,\,\,\,\,\,\,\,\,\,\,}^\gamma}
\newcommand{\sXlmq}{\,\;X_{\ell,m\!\!\!\!\!\!\!\!\!\!\!\!\!\!\!\!\!\!\!\!s\,\,\,\,\,\,\,\,\,\,\,\,\,\,\,\,}^{{\color{white} \gamma}}}
\newcommand{\sbllpmq}{\,\;b_{\ell,\ell+1\!\!\!\!\!\!\!\!\!\!\!\!\!\!\!\!\!\!\!\!\!s\,\,\,\,\,\,\,\,\,\,\,\,\,\,\,\,\,\,\,}^m}
\newcommand{\sbllmmq}{\,\;b_{\ell,\ell-1\!\!\!\!\!\!\!\!\!\!\!\!\!\!\!\!\!\!\!\!\!s\,\,\,\,\,\,\,\,\,\,\,\,\,\,\,\,\,\,\,}^m}
\newcommand{\sYlpmq}{\,\;Y_{\ell+1,m\!\!\!\!\!\!\!\!\!\!\!\!\!\!\!\!\!\!\!\!\!\!\!\!\!s\,\,\,\,\,\,\,\,\,\,\,\,\,\,\,\,\,\,\,\,\,\,}^{{\color{white} \gamma}}}
\newcommand{\sYlmmq}{\,\;Y_{\ell-1,m\!\!\!\!\!\!\!\!\!\!\!\!\!\!\!\!\!\!\!\!\!\!\!\!\!s\,\,\,\,\,\,\,\,\,\,\,\,\,\,\,\,\,\,\,\,\,\,}^{{\color{white} \gamma}}}
\newcommand{\sXlpmq}{\,\;X_{\ell+1,m\!\!\!\!\!\!\!\!\!\!\!\!\!\!\!\!\!\!\!\!\!\!\!\!\!\!\!\!s\,\,\,\,\,\,\,\,\,\,\,\,\,\,\,\,\,\,\,\,\,\,\,\,}^{{\color{white} \gamma}}}
\newcommand{\sXlmmq}{\,\;X_{\ell-1,m\!\!\!\!\!\!\!\!\!\!\!\!\!\!\!\!\!\!\!\!\!\!\!\!\!\!\!\!s\,\,\,\,\,\,\,\,\,\,\,\,\,\,\,\,\,\,\,\,\,\,\,\,}^{{\color{white} \gamma}}}
\newcommand{\sYlmprimeq}{\,\;Y_{\ell,m\!\!\!\!\!\!\!\!\!\!\!\!\!\!\!\!\!s\,\,\,\,\,\,\,\,\,\,\,\,\,\,}^{{\prime {\color{white} \gamma}}}}
\newcommand{\sXlmprimeq}{\,\;X_{\ell,m\!\!\!\!\!\!\!\!\!\!\!\!\!\!\!\!\!\!\!\!s\,\,\,\,\,\,\,\,\,\,\,\,\,\,\,\,}^{{\prime {\color{white} \gamma}}}}
\newcommand{\sYlpmprimeq}{\,\;Y_{\ell+1,m\!\!\!\!\!\!\!\!\!\!\!\!\!\!\!\!\!\!\!\!\!\!\!\!\!s\,\,\,\,\,\,\,\,\,\,\,\,\,\,\,\,\,\,\,\,\,\,}^{{\prime {\color{white} \gamma}}}}
\newcommand{\sXlpmprimeq}{\,\;X_{\ell+1,m\!\!\!\!\!\!\!\!\!\!\!\!\!\!\!\!\!\!\!\!\!\!\!\!\!\!\!\!s\,\,\,\,\,\,\,\,\,\,\,\,\,\,\,\,\,\,\,\,\,\,\,\,}^{{\prime {\color{white} \gamma}}}}
\newcommand{\sYlmmprimeq}{\,\;Y_{\ell-1,m\!\!\!\!\!\!\!\!\!\!\!\!\!\!\!\!\!\!\!\!\!\!\!\!\!s\,\,\,\,\,\,\,\,\,\,\,\,\,\,\,\,\,\,\,\,\,\,}^{{\prime {\color{white} \gamma}}}}
\newcommand{\sXlmmprimeq}{\,\;X_{\ell-1,m\!\!\!\!\!\!\!\!\!\!\!\!\!\!\!\!\!\!\!\!\!\!\!\!\!\!\!\!s\,\,\,\,\,\,\,\,\,\,\,\,\,\,\,\,\,\,\,\,\,\,\,\,}^{{\prime {\color{white} \gamma}}}}
\newcommand{\spSlmgq}{\,\;S_{\ell,m\!\!\!\!\!\!\!\!\!\!\!\!\!\!\!\!\!\!\!\!\!\!\!\!\!s+1\,\,\,\,\,\,\,\,\,\,\,\,\,\,\,\,}^\gamma}
\newcommand{\spTlmgq}{\,\;T_{\ell,m\!\!\!\!\!\!\!\!\!\!\!\!\!\!\!\!\!\!\!\!\!\!\!\!\!s+1\,\,\,\,\,\,\,\,\,\,\,\,\,\,\,\,}^\gamma}
\newcommand{\spbllpmq}{\,\;b_{\ell,\ell+1\!\!\!\!\!\!\!\!\!\!\!\!\!\!\!\!\!\!\!\!\!\!\!\!\!\!\!\!s+1\,\,\,\,\,\,\,\,\,\,\,\,\,\,\,\,\,\,\,}^m}
\newcommand{\spbllmmq}{\,\;b_{\ell,\ell-1\!\!\!\!\!\!\!\!\!\!\!\!\!\!\!\!\!\!\!\!\!\!\!\!\!\!\!\!s+1\,\,\,\,\,\,\,\,\,\,\,\,\,\,\,\,\,\,\,}^m}
\newcommand{\sSlmgprimeq}{\,\;S_{\ell,m\!\!\!\!\!\!\!\!\!\!\!\!\!\!\!\!\!\!s\,\,\,\,\,\,\,\,\,\,\,\,\,\,\,}^{\gamma\prime}}
\newcommand{\sTlmgprimeq}{\,\;T_{\ell,m\!\!\!\!\!\!\!\!\!\!\!\!\!\!\!\!\!\!s\,\,\,\,\,\,\,\,\,\,\,\,\,\,\,}^{\gamma\prime}}
\newcommand{\spYlmq}{\,\;Y_{\ell,m\!\!\!\!\!\!\!\!\!\!\!\!\!\!\!\!\!\!\!\!\!\!\!\!s+1\,\,\,\,\,\,\,\,\,\,\,\,\,\,}^{{\color{white} \gamma}}}
\newcommand{\spXlmq}{\,\;X_{\ell,m\!\!\!\!\!\!\!\!\!\!\!\!\!\!\!\!\!\!\!\!\!\!\!\!\!\!\!s+1\,\,\,\,\,\,\,\,\,\,\,\,\,\,\,\,}^{{\color{white} \gamma}}}
\newcommand{\spYlpmq}{\,\;Y_{\ell+1,m\!\!\!\!\!\!\!\!\!\!\!\!\!\!\!\!\!\!\!\!\!\!\!\!\!\!\!\!\!\!\!\!s+1\,\,\,\,\,\,\,\,\,\,\,\,\,\,\,\,\,\,\,\,\,\,}^{{\color{white} \gamma}}}
\newcommand{\spXlpmq}{\,\;X_{\ell+1,m\!\!\!\!\!\!\!\!\!\!\!\!\!\!\!\!\!\!\!\!\!\!\!\!\!\!\!\!\!\!\!\!\!\!s+1\,\,\,\,\,\,\,\,\,\,\,\,\,\,\,\,\,\,\,\,\,\,\,}^{{\color{white} \gamma}}}
\newcommand{\spYlmmq}{\,\;Y_{\ell-1,m\!\!\!\!\!\!\!\!\!\!\!\!\!\!\!\!\!\!\!\!\!\!\!\!\!\!\!\!\!\!\!\!s+1\,\,\,\,\,\,\,\,\,\,\,\,\,\,\,\,\,\,\,\,\,\,}^{{\color{white} \gamma}}}
\newcommand{\spXlmmq}{\,\;X_{\ell-1,m\!\!\!\!\!\!\!\!\!\!\!\!\!\!\!\!\!\!\!\!\!\!\!\!\!\!\!\!\!\!\!\!\!\!s+1\,\,\,\,\,\,\,\,\,\,\,\,\,\,\,\,\,\,\,\,\,\,\,}^{{\color{white} \gamma}}}
\newcommand{\sbetalmgq}{\,\;\beta_{\ell,m\!\!\!\!\!\!\!\!\!\!\!\!\!\!\!\!\!\!s\,\,\,\,\,\,\,\,\,\,\,\,\,\,\,}^\gamma}
\newcommand{\salphalmgq}{\,\;\alpha_{\ell,m\!\!\!\!\!\!\!\!\!\!\!\!\!\!\!\!\!\!s\,\,\,\,\,\,\,\,\,\,\,\,\,\,\,}^\gamma}
\newcommand{\sbtildellmmq}{\,\;\tilde{b}_{\ell,\ell-1\!\!\!\!\!\!\!\!\!\!\!\!\!\!\!\!\!\!\!\!\!s\,\,\,\,\,\,\,\,\,\,\,\,\,\,\,\,\,\,\,}^m}
\newcommand{\sXtildelmmq}{\,\;\tilde{X}_{\ell-1,m\!\!\!\!\!\!\!\!\!\!\!\!\!\!\!\!\!\!\!\!\!\!\!\!\!\!\!\!s\,\,\,\,\,\,\,\,\,\,\,\,\,\,\,\,\,\,\,\,\,\,\,\,}^{{\color{white} \gamma}}}

\begin{document}

\title{Raising and Lowering operators of spin-weighted spheroidal harmonics}
%Extraction of very high precision linear-in-mass-ratio post-Newtonian parameters from self-force computations for circular orbits in Schwarzschild spacetime}

\author{Abhay G. Shah} 
\email{a.g.shah@soton.ac.uk}
\affiliation{School of Mathematics, University of Southampton, Southampton SO17 1BJ, United Kingdom}

\author{Bernard F. Whiting} 
\email{bernard@phys.ufl.edu}
\affiliation{Department of Physics, P.O. Box 118440, University of Florida, Gainesville, Florida 32611-8440, USA}
\affiliation{$\mathcal{G}\mathbb{R}\varepsilon{\mathbb{C}}\mathcal{O}$,
   Institut d'Astrophysique de Paris --- UMR 7095 du CNRS,
   \\ Universit\'e Pierre \& Marie Curie, 98\textsuperscript{bis}
   boulevard Arago, 75014 Paris, France}

%\date{2012}
%\pacs{04.30.Db, 04.25.Nx, 04.70.Bw}

\begin{abstract}
%%%%%%%%%%%%%
% Abstract comes here %
%%%%%%%%%%%%%
Differential operators for raising and lowering angular momentum for spherical harmonics are used widely in many branches of physics.  Less well known are raising and lowering operators for both spin and the azimuthal component of angular momentum\cite{Goldberg:1966uu}.
In this paper we generalize the spin-raising and lowering operators of spin-weighted spherical harmonics to operators linear-in-$\gamma$ for spin-weighted spheroidal harmonics, where $\gamma$ is an additional parameter present in the second order ordinary differential equation governing these harmonics.  Constructing these operators has required using all the $\ell$-, $s$- and $m$-raising and lowering operators (and various combinations of them) for spin-weighted spherical harmonics, which have been calculated and shown explicitly in this paper.  Following a well-defined procedure, the operators given could be generalized to higher powers in $\gamma$.
\end{abstract}

\maketitle

\section{Introduction}
Spin-weighted spherical harmonics, $\sYlmq$, occur in many areas of physics -- from quantum mechanics to geophysics.  By contrast, spin-weighted spheroidal harmonics, $\sSlmgq$, are much less well known, but they arise \emph{naturally} in general relativity, for any analysis of the angular dependence of propagating fields on rotating, Kerr black hole space-time backgrounds \cite{Teukolsky:1972my,Teukolsky:1973ha}, and are most studied in the differential equations governing scalar, linear electromagnetic and gravitational perturbations. 
When the spin, $s$, of the propagating field is zero, these angular eigenfunctions become the oblate (scalar) spheroidal harmonics \cite{Flammer57}. Apart from their application in astrophysics, ordinary spheroidal harmonics ($s=0$ of $\sSlmgq$), $S_{\ell,m}^\gamma$, are also used to study molecular physics; for example, they occur in equations describing the hydrogen molecule ion or an electron in a dipole field \cite{Leaver}. Non-zero spin harmonics, the$\sSlmgq$, have been used when studying canonical quantization of electromagnetic field on a Kerr background \cite{CasalsOttewill}, and Hawking radiation from higher dimensional rotating black holes \cite {CKW} (also see references in Berti \etal\cite{BCC} for further applications). Only when the black hole is spherically symmetric, do the full angular eigenfunctions reduce to the spin-weighted spherical harmonics, well known in other areas of Physics.  

Though perhaps not very well known, it is actually possible to identify operators for raising and lowering the spin of spin-weighted spheroidal harmonics. 
%\textcolor{magenta}{A question that motivated this work was that just like spin-weighted spherical harmonics are built from ordinary spherical harmonics by repeatedly applying spin-raising or lowering operators, is it possible to build spin-weighted spheroidal harmonics from ordinary spheroidal harmonics?} 
In this work we will focus, for the first time, on describing such operators, \ie for raising and lowering the spin index of the spin-weighted spheroidal harmonics.  We explain how these operators can be described exactly, and then will show explicitly how they can be constructed to first order in the parameter perturbing away from the spin-weighted spherical harmonics, and lay the groundwork for extending the procedure to higher order.  For simplicity, we will generally assume an unwritten factor of $e^{im\phi}$ throughout, and shall concentrate primarily on the $\theta$-dependence, since the azimuthal eigen-equation is rather trivial.

%s represents the spin of the field, $m$ its azimuthal eigenvalue

Spin-weighted spheroidal harmonics satisfy the angular part of Teukolsky's master equation:
\begin{align} \label{sSlmg_eqn}
\ssquarelmgq\, z=\frac{1}{\sin\theta}\frac{d}{d\theta}\left( \sin\theta \frac{d z}{d\theta} \right) - \left( \frac{m^2+s^2+2\,m\,s\cos\theta}{\sin^2\theta} -\g^2\cos^2\theta+2s\gamma\cos\theta- \sElmgq\right) z = 0
\end{align}
where $s$ is the spin weight of the harmonic, and $\sElmgq$ represents a discrete set of eigenvalues which allow a solution to be regular in the whole interval $-1\le\cos\theta\le1$.  In the limit $\g \rightarrow 0$, $\sElmgq$ is just $\ell(\ell+1)$. As with any %$2^\textrm{nd}$ 
second order differential equation, Eq. (\ref{sSlmg_eqn}) has two linearly independent solutions, one of which, $\sSlmgq$ is required to be regular everywhere on the sphere and is used for describing scalar, (masless) neutrino, electromagnetic and gravitational perturbations. In the limit $\g\rightarrow0$, these harmonics are the spin-weighted spherical harmonics, $\sYlmq$. In the limit $s\rightarrow0$, $\sSlmgq$ are the ordinary spheroidal harmonics, $S_{\ell,m}^\g$, and $\sYlm$ are the ordinary spherical harmonics, $Y_{\ell,m}$.
The $\sYlmq$ appear as a solution to the equation
\begin{align} \label{sYlmeqn}
\frac{1}{\sin\theta}\frac{d}{d\theta}\left( \sin\theta \frac{d z}{d\theta} \right) - \left( \frac{m^2+s^2+2\,ms\cos\theta}{\sin^2\theta} - \ell(\ell+1)\right) z = 0.
\end{align}
To build $\sYlmq$ for non-zero $s$, one repeatedly applies spin raising and lowering operators on ordinary spherical harmonics, with eigenfunctions of positive and negative values of spin-weight being computed separately:
\begin{align}
\,_sY_{\ell,m} &= \eth_{s-1}\eth_{s-2}\cdots\eth_1\eth_0 \,\,\,_0Y_{\ell,m},  \nonumber \\
\,_{-|s|}Y_{\ell,m} &= \bar{\eth}_{-|s|+1}\bar{\eth}_{-|s|+2}\cdots\bar{\eth}_{-1}\bar{\eth}_0 \,\,\,_0Y_{\ell,m},
\end{align}
where $s\ge0$, and for all $s$ we have the definitions:
\begin{align}
\eth_s &= \frac{-(\partial_\theta - m \csc\theta - s\cot\theta)}{\sqrt{(\ell-s)(\ell+s+1)}}, \nonumber \\
\bar{\eth}_s &= \frac{(\partial_\theta + m \csc\theta + s\cot\theta)}{\sqrt{(\ell+s)(\ell-s+1)}}, \nonumber \\
\,_0Y_{\ell,m} &= \sqrt{ \frac{ (2\ell+1)(\ell-m)! } { 4\pi(\ell+m)! } } P_\ell^m (\cos\theta) e^{i m \phi}.
\end{align}
Here $\eth_s$ is the raising operator acting on $\,_sY_{\ell,m}$, $\bar{\eth}_s$ is the lowering operator acting on $\,_{-s}Y_{\ell,m}$, and $P_\ell^m$ are the associated Legendre functions. $\eth$ and $\bar{\eth}$ have been given a subscript here to show which spin-weighted quantities they act on. 
For each $s$, the $\sYlm$ are complete and orthogonal functions on the 2-sphere, and are related to the Wigner D-rotation matrices by
\beq
\sYlm (\theta,\phi) = \sqrt{\frac{2\ell+1}{4\pi}} \, D^\ell_{-s,m}(\phi,\theta,0).
\eeq

%To continue ahead we will also need to define spin-weighted spherical harmonics of \emph{type-2}. We do so by using associated Legendre functions, $Q_\ell^m$.
%We first define \emph{spherical harmonics of type-2} by 
%\begin{align}
%\,_0X_{\ell,m} &:= \sqrt{ \frac{ (2\ell+1)(\ell-m)! } { 4\pi(\ell+m)! } } Q_\ell^m (\cos\theta) e^{im\phi}.
%\end{align}
%Spin-weighted spherical harmonics of type-2 can be similarly defined as
%\begin{align}
%\,_sX_{\ell,m} &:= \eth_{s-1}\eth_{s-2}\cdots\eth_1\eth_0 \,\,\,_0X_{\ell,m},  \nonumber \\
%\,_{-|s|}X_{\ell,m} &:= \bar{\eth}_{-|s|+1}\bar{\eth}_{-|s|+2}\cdots\bar{\eth}_{-1}\bar{\eth}_0 \,\,\,_0X_{\ell,m}.
%\end{align}

Unlike the way $\sYlmq$ are calculated using raising and lowering operators, spin-weighted spheroidal harmonics are usually calculated as a sum over $\sYlmq$ (see Press and Teukolsky\cite{Press:1973zz}) or as a sum over Jacobi polynomials \cite{Fackerell:JMP1977}. In this paper we work on generalizing $\eth_s$ and $\bar{\eth}_s$ to operators that raise and lower the spin-weight of spheroidal harmonics, $\sSlmgq$. That is, if $z=\sSlmgq$ and $y=\,\,\,\,\spmSlmgq$ are solutions to two different differential versions of Eq. (\ref{sSlmg_eqn}), one being $\ssquarelmgq\,\, z=0$ and the other being $\,\,\,\,\spmsquarelmgq\,\, y=0$, then we will find a relation of the form:
\beq
y=\alpha \, z + \beta\, \partial_\theta z,\label{abt}
\eeq
to linear order in $\gamma$, between the solutions ($y$ and $z$) of these equations.

The paper is organized as follows. In Section II, we summarize Whiting's earlier work \cite{Whiting:1984sf} on finding relations between solutions of two differential equations. In Section III, we use this work to calculate the different $\ell$-, $s$- and $m$-raising and lowering relations of $\sYlmq$. In Section IV, we build the linear-in-$\g$ $s$-raising and lowering operators for $\sSlmgq$.

%The spin-raising and lowering operators of spin-weighted spherical harmonics, which are generally represented by $\eth$ and $\bar{\eth}$, respectively, have been known for a long time. One builds positive-s and negative-s valued $\sYlm$ by successively applying these $\eth$ ad $\bar{\eth}$ operators on ordinary spherical harmonics, $Y_{\ell,m}$ or $\,_0Y_{\ell,m}$.

\section{Earlier work on relating solutions of two differential equations} \label{sec2}

Relations of the general form which we seek have been studied previously by one of us \cite{Whiting:1984sf} and were extensively used in \cite{Whiting:1988vc} to show mode stability for the perturbations being discussed here.  We now give a brief, and slightly more general, introduction, while more complete details can be found in \cite{Whiting:1984sf}.  Thus, we suppose that $y(x)$ and $z(x)$ satisfy 
\beq
y''+p\,y'+q\,y=0\quad{\rm and}\quad z''+P\,z'+Q\,z=0,%\quad{\rm where}\quad y=\alpha z+\beta z' 
\label{2nd}
\eeq
in which $'=d/dx$, and seek conditions that $\alpha$ and $\beta$ must satisfy in order that
\beq
y=\alpha \, z+\beta \, z'\label{ab}
\eeq
should hold.  More specifically, since each of Eq. (\ref{2nd}) is second order, two linearly independent solutions exist, say $(y_1, y_2)$ and $(z_1, z_2)$ respectively, and we will actually demand that the mapping (\ref{ab}) applies more fully, so that:
\bal
\begin{split}
y_1&=\alpha \, z_1 + \beta \, z'_1,\\
y_2&=\alpha \, z_2 + \beta \, z'_2.
\end{split}
\label{y12}
\end{align}
That is, every solution for $y$ will map to a solution for $z$.  Defining the relevant Wronskians by:
\beq
W(y_1,y_2)\equiv W_y=y_1\, y_2'-y_1'\, y_2=C_ye^{-\int p \,dx}\quad{\rm and}\quad W(z_1,z_2)\equiv W_z=z_1\, z_2'-z_1'\, z_2=C_ze^{-\int P \, dx},\label{Wronskians}
\eeq
where $C_y$ and $C_z$ are constants, we can invert Eqs (\ref{y12}) to find $\alpha$ and $\beta$:
\bal
\begin{split}
\alpha&=\frac{1}{W_z}\left(y_1\,z_2'-y_2\,z_1'\right),\\
\beta&=-\frac{1}{W_z}\left(y_1\,z_2-y_2\,z_1\right).
\end{split}
\label{abdef}
\end{align}
Clearly, $\alpha$ and $\beta$ are determined entirely by the solutions they map between.  Differentiating (\ref{ab}) once and using Eq. (\ref{2nd}) for $z$ we find:
\beq
y'=(\alpha'-\beta \,Q)z+(\alpha+\beta'-\beta \,P)z'.\label{yp}
\eeq
Eqs (\ref{ab}) and (\ref{yp}) together can be inverted to give $z$ and $z'$ in terms of $y$ and $y'$.  For this we will also need to define:
\beq
k=\alpha(\alpha+\beta'-\beta \, P)-\beta(\alpha'-\beta \, Q)=\frac{W_y}{W_z}.\label{eqk}
\eeq
Then
\bal
\begin{split}
z&=\frac{1}{k}\left((\alpha+\beta'-\beta \, P)y-\beta \, y' \right),\\
z'&=\frac{1}{k}\left(-(\alpha'-\beta \, Q)y+\alpha \, y' \right).
\end{split}
\label{zzp}
\end{align}
Further differentiating Eq. (\ref{yp}), and using both Eqs (\ref{2nd}), we can deduce:
\bal
0&=\left(\alpha''+p\, \alpha'+(q-Q)\alpha-2Q\, \beta'-Q'\, \beta-(p-P)Q\, \beta\right)z\nonumber\\
&\eq+\left(2\alpha'+(p-P)\alpha+\beta''+(p-2P)\beta'+(q-Q)\beta-P'\, \beta-(p-P)P\,\beta\right)z',
\end{align}
in which each coefficient must separately be zero because of Eqs (\ref{y12}).  Thus:
\bal
\begin{split}
\alpha''+p \, \alpha'+(q-Q)\alpha-2Q\, \beta'-Q'\, \beta-(p-P)Q\, \beta&=0\\
2\alpha'+(p-P)\alpha+\beta''+(p-2P)\beta'+(q-Q)\beta-P'\, \beta-(p-P)P\, \beta&=0.
\end{split}
\label{abrel}
\end{align}
With the appropriate combination of these, we can now show constructively that:
\beq
k'=(P-p)k,\label{kp}
\eeq
as already follows from Eqs (\ref{Wronskians}) and (\ref{eqk}) above.  In the application we have in mind, $P=p$, so that $k=\rm{const}$.  We could also check the integrability of Eqs (\ref{zzp}) which, with Eq. (\ref{kp}) and some algebra, yields the second of Eqs (\ref{abrel}).  

Finally we note that the operators in Eq. (\ref{2nd}) can be written as:
\bal
\begin{split}
\partial_{xx}+P\,\partial_x+Q&=\left(\partial_x-\frac{\alpha}{\beta}+P\right)\left(\partial_x+\frac{\alpha}{\beta}\right)+\frac{k}{\beta^2},\\
\partial_{xx}+p\,\partial_x+q&=\left(\partial_x+\frac{\alpha+\beta'}{\beta}+p-P\right)\left(\partial_x-\frac{\alpha+\beta'}{\beta}+P\right)+\frac{k}{\beta^2}.\\
\end{split}
\label{oxy}
\end{align}
in which the first order operators are effectively intertwined.

\section{Spin-weighted spherical harmonics} \label{sec3}

Let us denote spin-weighted spherical harmonics of type-1 and type-2 by $\,_sY_{\ell,m}$ and $\,_sX_{\ell,m}$, being two linearly independent solutions of the following equation,
\begin{align} \label{sYlmeqn}
\frac{1}{\sin\theta}\frac{d}{d\theta}\left( \sin\theta \frac{d z}{d\theta} \right) - \left( \frac{m^2+s^2+2\,m\,s\cos\theta}{\sin^2\theta} - \ell(\ell+1)\right) z = 0,
\end{align}
where, in the notation of section \ref{sec2},
\begin{align}
z_1 &= \sYlm,  \, \textrm{ and} \nonumber\\
z_2 &= \sXlm.
\end{align}

To build the harmonics of non-zero spin, we begin with spin-weight zero ordinary spherical harmonics (suppressing $e^{im\phi}$):
\begin{align}
\,_0Y_{\ell,m} &= \sqrt{ \frac{ (2\ell+1)(\ell-m)! } { 4\pi(\ell+m)! } } P_\ell^m (\cos\theta), \nonumber \\
\,_0X_{\ell,m} &= \sqrt{ \frac{ (2\ell+1)(\ell-m)! } { 4\pi(\ell+m)! } } Q_\ell^m (\cos\theta),
\end{align}
and apply further $s$-raising and $s$-lowering operators to generate arbitrary spin-weighted spherical harmonics:
\begin{align}
\eth_s &= \frac{-(\partial_\theta - m \csc\theta - s\cot\theta)}{\sqrt{(\ell-s)(\ell+s+1)}}, \nonumber \\
\bar{\eth}_s &= \frac{(\partial_\theta + m \csc\theta + s\cot\theta)}{\sqrt{(\ell+s)(\ell-s+1)}}.
\end{align}

Therefore,
\begin{align}
\,_sY_{\ell,m} &= \eth_{s-1}\eth_{s-2}\cdots\eth_1\eth_0 \,\,\,_0Y_{\ell,m},  \nonumber \\
\,_{-|s|}Y_{\ell,m} &= \bar{\eth}_{-|s|+1}\bar{\eth}_{-|s|+2}\cdots\bar{\eth}_{-1}\bar{\eth}_0 \,\,\,_0Y_{\ell,m},
\end{align}
and same holds for $\,_sX_{\ell,m}$.

From Section \ref{sec2} we know that to find relations between solutions of equations
\begin{align}
z^{\prime\prime} + P\,z' +Q\,z=0,\qquad \qquad y^{\prime\prime} + p\,y'+q\,y=0
\end{align}
we need to calculate
\begin{align}
\beta &= \frac{z_1 \, y_2 - z_2 \, y_1}{W(z_1,z_2)}, \, \textrm{ and} \nonumber \\
\alpha &= \frac{-z_1^\prime \, y_2 + z_2^\prime \, y_1}{W(z_1,z_2)},
\end{align}
where $W(z_1,z_2)$ is defined in (\ref{Wronskians}).  % = z_1 z_2^\prime - z_2 z_1^\prime$.
One then has
\begin{align}
y_i = \beta \, z_i^\prime + \alpha \, z_i .
\end{align}

Finding $\beta$'s and $\alpha$'s for various relations, we get
\begin{align} %\allowdisplaybreaks
\,_{s+1}Y_{\ell,m} &= \frac{-1}{\sqrt{(\ell-s)(\ell+s+1)}} \left( \partial_\theta - \frac{m}{\sin\theta} - s\cot\theta \right) \,_sY_{\ell,m} \label{eqsp}\\
\,_{s-1}Y_{\ell,m} &= \frac{1}{\sqrt{(\ell+s)(\ell-s+1)}} \left( \partial_\theta + \frac{m}{\sin\theta} + s\cot\theta \right) \,_sY_{\ell,m} \label{eqsm}\\ 
\,_{s}Y_{\ell+1,m} &= \sqrt{\frac{2\ell+3}{2\ell+1}}\frac{(\ell+1)\sin\theta}{\sqrt{ \left[(\ell+1)^2-m^2\right]\left[(\ell+1)^2-s^2\right] }} \left( \partial_\theta + \frac{m \, s}{(\ell+1)\sin\theta} + (\ell+1)\cot\theta \right) \,_sY_{\ell,m} \label{eqlp}\\ 
\,_{s}Y_{\ell-1,m} &= \sqrt{\frac{2\ell-1}{2\ell+1}}\frac{-\,\ell\sin\theta}{\sqrt{ \left(\ell^2-m^2\right)\left(\ell^2-s^2\right) }} \left( \partial_\theta - \frac{m\, s}{\ell\sin\theta} - \ell\cot\theta \right) \,_sY_{\ell,m} \label{eqlm}\\ %\displaybreak
\,_{s}Y_{\ell,m+1} &= \frac{1}{\sqrt{(\ell-m)(\ell+m+1)}} \left( \partial_\theta - \frac{s}{\sin\theta} - m\cot\theta \right) \,_sY_{\ell,m} \label{eqmp}\\ 
\,_{s}Y_{\ell,m-1} &= \frac{-1}{\sqrt{(\ell+m)(\ell-m+1)}} \left( \partial_\theta + \frac{s}{\sin\theta} + m\cot\theta \right) \,_sY_{\ell,m} \label{eqmm}
\end{align}
The above six relations are equivalent to Gauss's relations for contiguous functions of the hypergeometric function, $\,_2F_1(a,b,c;z)$. By equating $\partial_\theta \,_sY_{\ell,m}$ of two different relations, one can get various numbers of recurrence relations. For example, by equating $\partial_\theta \,_sY_{\ell,m}$ in Eqs (\ref{eqsp}) and (\ref{eqlm}), one gets a relations between $\,_sY_{\ell,m}$, $\,_{s+1}Y_{\ell,m}$ and $\,_sY_{\ell-1,m}$. By repeated application of this procedure, various relations can be formed between different $\,_{s+i}Y_{\ell+j,m+k}$ (where $i,j,k$ are integers).

Finally, as an example of \Deqn{oxy} we show:
\begin{align} \allowdisplaybreaks
\eq{}&\;\partial_{\theta\theta}+\cot{\theta}\partial_\theta-\frac{m^2+s^2+2m\,s\cos\theta}{\sin^2\theta}+\ell(\ell+1)\\
={}&\left(\partial_\theta+\frac{m}{\sin\theta}+(s+1)\cot\theta\right)\left(\partial_\theta-\frac{m}{\sin\theta}-s\cot\theta\right)+(\ell-s)(\ell+s+1)\\
={}&\left(\partial_\theta-\frac{m}{\sin\theta}-(s-1)\cot\theta\right)\left(\partial_\theta+\frac{m}{\sin\theta}+s\cot\theta\right)+(\ell+s)(\ell-s+1)\\
={}&\left(\partial_\theta-\frac{m \, s}{(\ell+1)\sin\theta}-\ell\cot\theta\right)\left(\partial_\theta+\frac{m\,s}{(\ell+1)\sin\theta}+(\ell+1)\cot\theta\right) \nonumber \\& +\frac{\left((\ell+1)^2-m^2\right)\left((\ell+1)^2-s^2\right)}{(\ell+1)^2\sin^2\theta} \\
={}&\left(\partial_\theta+\frac{m\,s}{\ell\sin\theta}+(\ell+1)\cot\theta\right)\left(\partial_\theta-\frac{m\,s}{\ell\sin\theta}-\ell\cot\theta\right)+\frac{\left(\ell^2-m^2\right)\left(\ell^2-s^2\right)}{\ell^2\sin^2\theta}  \\
={}&\left(\partial_\theta+\frac{s}{\sin\theta}+(m+1)\cot\theta\right)\left(\partial_\theta-\frac{s}{\sin\theta}-m\cot\theta\right)+(\ell-m)(\ell+m+1)\displaybreak \\
={}&\left(\partial_\theta-\frac{s}{\sin\theta}-(m-1)\cot\theta\right)\left(\partial_\theta+\frac{s}{\sin\theta}+m\cot\theta\right)+(\ell+m)(\ell-m+1).
\end{align}

\section{Spin-weighted spheroidal harmonics}
\label{sec4}
Let us denote the spin-weighted spheroidal harmonics of type-1 and type-2 by $\sSlmgq$ and $\sTlmgq$, being two linearly independent solutions of the following equation,
\begin{align} \label{sSlmg_eqn_lin}
\frac{1}{\sin\theta}\frac{d}{d\theta}\left( \sin\theta \frac{d z}{d\theta} \right) - \left( \frac{m^2+s^2+2\,m\,s\cos\theta}{\sin^2\theta} -\g^2\cos^2\theta+2s\,\gamma\cos\theta- \sElmgq \right) z = 0.
\end{align}
To linear-in-$\g$, expressions for $\sSlmgq$ and $\sElmgq$ have been given by Press and Teukolsky \cite{Press:1973zz}.  We write:
\begin{align}
\sSlmgq &= \sYlmq + \g \, \sbllpmq \sYlpmq + \g \, \sbllmmq \sYlmmq + O(\g^2), \,\textrm{and}\nonumber \\
\sTlmgq &= \sXlmq + \g \, \sbllpmq \sXlpmq + \g \, \sbllmmq \sXlmmq + O(\g^2). \nonumber \\
\end{align}
We first look at the Wronskian to $O(\g)$,
\begin{align}
&W(\sSlmgq,\sTlmgq) = \left( \sYlmq\,\,\sXlmprimeq - \sXlmq\,\,\sYlmprimeq \right) + \g\, \sbllpmq\left( -\sYlpmprimeq \,\,\sXlmq + \sXlpmprimeq \,\,\sYlmq \right) \nonumber \\ &\quad+ \g \,\,\sbllpmq\left( -\sYlmprimeq \,\,\sXlpmq + \sXlmprimeq \,\,\sYlpmq \right) + \g \, \sbllmmq \left( -\sYlmmprimeq \,\, \sXlmq + \sXlmmprimeq \,\, \sYlmq \right) \nonumber \\ &\quad+ \g \,\, \sbllmmq \left( -\sYlmprimeq \,\, \sXlmmq + \sXlmprimeq \,\, \sYlmmq \right)  + O(\g^2).
\end{align}

In the above equation, the expression in the first parenthesis is a Wronskian which we denote by $\,_sw_{\ell,m}$, the expression in the second parenthesis is $\alpha \,_sw_{\ell+1,m}$ of $(\sYlpm \rightarrow \sYlm)$, the expression in the third parenthesis is $\alpha \,_sw_{\ell,m}$ of $(\sYlm \rightarrow \sYlpm)$, the expression in the fourth parenthesis is $\alpha \,_sw_{\ell-1,m}$ of $(\sYlmm\rightarrow\sYlm)$, and the expression in the fifth parenthesis is $\alpha \,_sw_{\ell,m}$ of $(\sYlm\rightarrow\sYlmm)$, all of which are known.
The analytical expression of $b$'s and $\alpha$'s are
\begin{align} \allowdisplaybreaks
&\sbllpmq = - \frac{ s\sqrt{\left[(\ell+1)^2-m^2\right]\left[(\ell+1)^2-s^2\right] } }{ (\ell+1)^2\sqrt{(2\ell+1)(2\ell+3)} },  \\
&\sbllmmq = \frac{ s\sqrt{(\ell^2-m^2)(\ell^2-s^2)} }{ \ell^2\sqrt{(2\ell-1)(2\ell+1)} },   \\
&\,_sw_{\ell,m} = W(\sYlm,\sXlm) = \frac{-(2\ell+1)\csc\theta}{4\pi}, \\
&\alpha_{(s,\ell+1,m)\rightarrow(s,\ell,m)} = \sqrt{\frac{2\ell+1}{2\ell+3}} \frac{(\ell+1)\sin\theta}{\sqrt{ \left[(\ell+1)^2-m^2\right]\left[(\ell+1)^2-s^2\right] }} \left( \frac{m\,s}{(\ell+1)\sin\theta} + \left(\ell+1\right)\cot\theta \right),  \displaybreak \\
&\alpha_{(s,\ell,m)\rightarrow(s,\ell+1,m)} = \sqrt{\frac{2\ell+3}{2\ell+1}} \frac{(\ell+1)\sin\theta}{\sqrt{ \left[(\ell+1)^2-m^2\right]\left[(\ell+1)^2-s^2\right] }} \left( \frac{m\,s}{(\ell+1)\sin\theta} + (\ell+1)\cot\theta \right),  \\
&\alpha_{(s,\ell-1,m)\rightarrow(s,\ell,m)} = \sqrt{\frac{2\ell+1}{2\ell-1}} \frac{\ell\sin\theta}{\sqrt{ \left(\ell^2-m^2\right)\left(\ell^2-s^2\right) }} \left( \frac{m\,s}{\ell\sin\theta} + \ell\cot\theta \right), \\
&\alpha_{(s,\ell,m)\rightarrow(s,\ell-1,m)} = \sqrt{\frac{2\ell-1}{2\ell+1}} \frac{\ell\sin\theta}{\sqrt{ \left(\ell^2-m^2\right)\left(\ell^2-s^2\right) }} \left( \frac{m\,s}{\ell\sin\theta} + \ell\cot\theta \right).  
\end{align}

Using these together, we get
\begin{align}
W(\sSlmgq,\sTlmgq) &= -\frac{2\ell+1}{4\pi\sin\theta} \left(1+\frac{2\,m\,s^2\g}{\ell^2(\ell+1)^2} \right) + O(\g^2).
\end{align}

We now wish to find the s-raising operator of $\sSlmgq$ to linear-order-in-$\g$. We start with finding the $\beta$ and $\alpha$ in the following relation
\begin{align}
y = \beta \, z^\prime + \alpha \, z + O(\g^2)
\end{align}
where $y =\,\,\,\, \spSlmgq,\,\,\,\,\,\,\, \spTlmgq\,$, and $z = \sSlmgq,\,\sTlmgq$. 

We first look at $\beta$:
\begin{align}
\beta &= \frac{\sSlmgq \,\,\,\,\,\,\, \spTlmgq - \sTlmgq \,\,\,\,\,\,\, \spSlmgq}{W(\sSlmgq,\sTlmgq)} \nonumber \\
&= -\frac{4\pi\sint}{2\ell+1}\left( 1 - \frac{2ms^2\g}{\ell^2(\ell+1)^2} \right) \left( \sSlmgq \,\,\,\,\,\,\, \spTlmgq - \sTlmgq \,\,\,\,\,\,\, \spSlmgq \right) + O(\g^2).
\end{align}

To calculate the above, we need
\begin{align} \label{beta}
\left( \sSlmgq \,\,\,\,\,\,\, \spTlmgq - \sTlmgq \,\,\,\,\,\,\, \spSlmgq \right) 
%=& \left( \tsYlm \, \tspXlm - \tsXlm \, \tspYlm \right) \nonumber \\ 
%&+ \g \sbllpm \left( \tsYlpm \, \tspXlm - \tsXlpm \, \tspYlm \right) \nonumber \\
%&+ \g \sbllmm \left( \tsYlmm \, \tspXlm - \tsXlmm \, \tspYlm \right) \nonumber \\
%&+ \g \spbllpm \left( \tsYlm \, \tspXlpm - \tsXlm \, \tspYlpm \right) \nonumber \\
%&+ \g \spbllmm \left( \tsYlm \, \tspXlmm - \tsXlm \, \tspYlmm \right) \nonumber \\
% 
=& \left( \sYlm \, \spXlm - \sXlm \, \spYlm \right) \nonumber \\ 
&+ \g \, \sbllpmq  \left(   \sYlpm  \,  \spXlm    -   \sXlpm   \,  \spYlm \right) \nonumber \\
&+ \g \, \sbllmmq  \left(  \sYlmm \,  \spXlm    -   \sXlmm  \,   \spYlm \right) \nonumber \\
&+ \g\,\,\,\,\,\, \spbllpmq  \left( \sYlm    \,  \spXlpm  -   \sXlm     \,   \spYlpm \right) \nonumber \\
&+ \g\,\,\,\,\,\, \spbllmmq  \left( \sYlm   \,  \spXlmm -   \sXlm     \,   \spYlmm \right). 
\end{align}

We next look at $\alpha$:
\begin{align}
\alpha &= \frac{- \sSlmgprimeq \,\,\,\,\,\,\, \spTlmgq + \sTlmgprimeq \,\,\,\,\,\,\, \spSlmgq}{W(\sSlmgq,\sTlmgq)} \nonumber \\
&= -\frac{4\pi\sint}{2\ell+1}\left( 1 - \frac{2\,m\,s^2\g}{\ell^2(\ell+1)^2} \right) \left( - \sSlmgprimeq \,\,\,\,\,\,\, \spTlmgq + \sTlmgprimeq \,\,\,\,\,\,\, \spSlmgq \right) + O(\g^2).
\end{align}

To calculate the above, we need
\begin{align} \label{alpha} \allowdisplaybreaks
\left( -\sSlmgprimeq \,\,\,\,\,\,\, \spTlmgq + \sTlmgprimeq \,\,\,\,\,\,\, \spSlmgq \right) 
%=& \left( \tsYlm \, \tspXlm - \tsXlm \, \tspYlm \right) \nonumber \\ 
%&+ \g \sbllpm \left( \tsYlpm \, \tspXlm - \tsXlpm \, \tspYlm \right) \nonumber \\
%&+ \g \sbllmm \left( \tsYlmm \, \tspXlm - \tsXlmm \, \tspYlm \right) \nonumber \\
%&+ \g \spbllpm \left( \tsYlm \, \tspXlpm - \tsXlm \, \tspYlpm \right) \nonumber \\
%&+ \g \spbllmm \left( \tsYlm \, \tspXlmm - \tsXlm \, \tspYlmm \right) \nonumber \\
% 
=& \left( -\sYlmprimeq \,\,\,\,\,\,\, \spXlmq + \sXlmprimeq \,\,\,\,\,\,\, \spYlmq \right) \nonumber \\ 
&+ \g \,\sbllpmq \left( - \sYlpmprimeq \,\,\,\,\,\,\, \spXlmq + \sXlpmprimeq \,\,\,\,\,\,\, \spYlmq \right) \nonumber  \\
&+ \g \,\sbllmmq  \left( - \sYlmmprimeq \,\,\,\,\,\,\, \spXlmq + \sXlmmprimeq \,\,\,\,\,\,\, \spYlmq \right) \nonumber \\
&+ \g \,\,\,\,\,\, \spbllpmq \left( - \sYlmprimeq \,\,\,\,\,\,\, \spXlpmq + \sXlmprimeq \,\,\,\,\,\,\, \spYlpmq \right) \nonumber   \\
&+ \g \,\,\,\,\,\, \spbllmmq  \left( - \sYlmprimeq \,\,\,\,\,\,\, \spXlmmq + \sXlmprimeq \,\,\,\,\,\,\, \spYlmmq \right). 
\end{align}

In Eqs (\ref{beta}, \ref{alpha}),
the expression in the first set of parenthesis are the $\beta\,_sw_{\ell,m}$ and $\alpha\,_sw_{\ell,m}$ of $\left( \sYlm \rightarrow \spYlm \right)$, 
the expression in the second set of parenthesis are the $\beta\,_sw_{\ell+1,m}$ and $\alpha\,_sw_{\ell+1,m}$ of $\left( \sYlpm\rightarrow \spYlm \right)$, 
the expression in the third set of parenthesis are the $\beta\,_sw_{\ell-1,m}$ and $\alpha\,_sw_{\ell-1,m}$ of $\left( \sYlmm \rightarrow \spYlm \right)$, 
the expression in the fourth set of parenthesis are the $\beta\,_sw_{\ell,m}$ and $\alpha\,_sw_{\ell,m}$ of $\left( \sYlm \rightarrow \spYlpm \right)$, and 
the expression in the fifth set of parenthesis are the $\beta\,_sw_{\ell,m}$ and $\alpha\,_sw_{\ell,m}$ of $\left( \sYlm \rightarrow \spYlmm \right)$.
To get the $\beta$'s and $\alpha$'s of the last four expressions, we will need the following relations,
\begin{align} \allowdisplaybreaks
%1
\spYlm =&  \sqrt{\frac{2\ell+1}{2\ell+3}} \frac{-(\ell+s+1)[m+(\ell+1)\cos\theta]}{ \sqrt{ \left[(\ell+1)^2-m^2\right] \left[(\ell+1)^2-s^2\right] (\ell-s) (\ell+s+1) } } \nonumber \\ & \times \left[ \partial_\theta - \frac{ m(\ell+s+1)\cot\theta }{ [m+(\ell+1)\cos\theta] } - \frac{ (\ell+1)^2(s-1)\cos\theta\cot\theta }{ (\ell+s+1)[m+(\ell+1)\cos\theta] } \right. \nonumber \\  & \left. - \frac{ \left[ \ell^2+s^2 +m^2(s+1) +\ell(m^2+s^2+2)+1 \right]\csc\theta }{ (\ell+s+1)[m+(\ell+1)\cos\theta] } \right. \nonumber \\ & \left. + \frac{ (\ell + 1)^2 (\ell + 2) \sint }{ (\ell+s+1)[m+(\ell+1)\cos\theta] } \right]\sYlpm \\ \nonumber \\ 
%2
\spYlm =& \sqrt{\frac{2\ell+1}{2\ell-1}} \frac{ -(\ell-s)(-m+\ell\cos\theta) }{ \sqrt{ \left(\ell^2-m^2\right) \left(\ell^2-s^2\right) (\ell-s) (\ell+s+1) } } \left[ \partial_\theta - \frac{ m(\ell-s)\cot\theta }{ (-m+\ell\cos\theta) } \right. \nonumber \\ & \left. - \frac{ \left(\ell^2-2m^2+2\ell s\right)\csc\theta }{ 2(-m+\ell\cos\theta) } + \frac{ \ell^2\cos2\theta\csc\theta }{ 2(-m+\ell\cos\theta) } \right]\sYlmm 
\\ \nonumber   \\
%3
\spYlpm =& \sqrt{ \frac{2\ell+3}{2\ell+1} } \frac{ -\sqrt{\ell-s+1}\,\,[-m+(\ell+1)\cos\theta] }{ \sqrt{ \left[(\ell+1)^2-m^2\right] \left[(\ell+1)^2-s^2\right](\ell+s+2) } } \nonumber  \\ & \left[ \partial_\theta - \frac{m (\ell - s + 1) \cot\theta}{ [-m + (\ell + 1) \cos\theta] } - \frac{ (\ell + 1)^2 (s - 1) \cos\theta \cot\theta }{ (\ell - s + 1) [-m + (\ell + 1) \cos\theta] } \right. \nonumber \\ &\left. + \frac{ \left[s^2 (\ell + 1) - (\ell - m^2 + 1) (\ell + 1) - m^2 s\right] \csc\theta }{ (\ell - s + 1) [-m + (\ell + 1) \cos\theta] } \right. \nonumber \displaybreak \\ &\left. - \frac{ \ell (\ell + 1)^2 \sin\theta }{ (\ell - s + 1) [-m + (\ell + 1) \cos\theta] } \right] \sYlm  \\ \nonumber \\
%4
\spYlmm =& \sqrt{\frac{2\ell-1}{2\ell+1}} \frac{ -(\ell + s) (m + \ell \cos\theta) }{ \sqrt{ (\ell^2 - m^2) (\ell^2 - s^2) (\ell + s) (\ell - s - 1) } } \left[ \partial_\theta  - \frac{ m (\ell + s) \cot\theta }{ (m + \ell\cos\theta) } \right. \nonumber \\ & \left. + \frac{ (\ell^2 - 2 m^2 - 2 \ell s) \csc\theta }{ 2(m + \ell\cos\theta) } - \frac{ \ell^2 \cos2\theta \csc\theta }{ 2(m + \ell\cos\theta) } \right] \sYlm
\end{align}

The results are as follows:
\begin{align} \allowdisplaybreaks
\beta_{(s,\ell,m)\rightarrow(s+1,\ell,m)} =& \frac{-1}{\sqrt{(\ell-s)(\ell+s+1)}}, \nonumber \\
\beta_{(s,\ell+1,m)\rightarrow(s+1,\ell,m)} =& \sqrt{\frac{2\ell+1}{2\ell+3}} \frac{-(\ell+s+1)[m+(\ell+1)\cos\theta]}{ \sqrt{ \left[(\ell+1)^2-m^2\right] \left[(\ell+1)^2-s^2\right] (\ell-s) (\ell+s+1) } }, \nonumber \\
\beta_{(s,\ell-1,m)\rightarrow(s+1,\ell,m)} =& \sqrt{\frac{2\ell+1}{2\ell-1}} \frac{ -(\ell-s)(-m+\ell\cos\theta) }{ \sqrt{ \left(\ell^2-m^2\right) \left(\ell^2-s^2\right) (\ell-s) (\ell+s+1) } }, \nonumber \\ 
\beta_{(s,\ell,m)\rightarrow(s+1,\ell+1,m)} =& \sqrt{ \frac{2\ell+3}{2\ell+1} } \frac{ -\sqrt{\ell-s+1}\,\,[-m+(\ell+1)\cos\theta] }{ \sqrt{ \left[(\ell+1)^2-m^2\right] \left[(\ell+1)^2-s^2\right](\ell+s+2) } }, \nonumber \\
\beta_{(s,\ell,m)\rightarrow(s+1,\ell-1,m)} =& \sqrt{\frac{2\ell-1}{2\ell+1}} \frac{ -(\ell + s) (m + \ell \cos\theta) }{ \sqrt{ (\ell^2 - m^2) (\ell^2 - s^2) (\ell + s) (\ell - s - 1) } }, \nonumber \\ 
\alpha_{(s,\ell,m)\rightarrow(s+1,\ell,m)} =& \frac{1}{\sqrt{(\ell-s)(\ell+s+1)}} \left[ \frac{m}{\sin\theta} + \left(s+\frac12\right)\cot\theta \right], \nonumber \\
%1
\alpha_{(s,\ell+1,m)\rightarrow(s+1,\ell,m)} =& \sqrt{\frac{2\ell+1}{2\ell+3}} \frac{-(\ell+s+1)[m+(\ell+1)\cos\theta]}{ \sqrt{ \left[(\ell+1)^2-m^2\right] \left[(\ell+1)^2-s^2\right] (\ell-s) (\ell+s+1) } } \nonumber \\ & \times \left[ - \frac{ m(\ell+s+1)\cot\theta }{ [m+(\ell+1)\cos\theta] } - \frac{ (\ell+1)^2(s-1)\cos\theta\cot\theta }{ (\ell+s+1)[m+(\ell+1)\cos\theta] } \right. \nonumber \\  & \left. - \frac{ \left[ \ell^2+s^2 +m^2(s+1) +\ell(m^2+s^2+2)+1 \right]\csc\theta }{ (\ell+s+1)[m+(\ell+1)\cos\theta] } \right. \nonumber \\ & \left. + \frac{ (\ell + 1)^2 (\ell + 2) \sint }{ (\ell+s+1)[m+(\ell+1)\cos\theta] } \right], \nonumber \\
%2
\alpha_{(s,\ell-1,m)\rightarrow(s+1,\ell,m)} =& \sqrt{\frac{2\ell+1}{2\ell-1}} \frac{ -(\ell-s)(-m+\ell\cos\theta) }{ \sqrt{ \left(\ell^2-m^2\right) \left(\ell^2-s^2\right) (\ell-s) (\ell+s+1) } } \left[ - \frac{ m(\ell-s)\cot\theta }{ (-m+\ell\cos\theta) } \right. \nonumber \\ & \left. - \frac{ \left(\ell^2-2m^2+2\ell s\right)\csc\theta }{ 2(-m+\ell\cos\theta) } + \frac{ \ell^2\cos2\theta\csc\theta }{ 2(-m+\ell\cos\theta) } \right], \nonumber \\
%3
\alpha_{(s,\ell,m)\rightarrow(s+1,\ell+1,m)} =& \sqrt{ \frac{2\ell+3}{2\ell+1} } \frac{ -\sqrt{\ell-s+1}\,\,[-m+(\ell+1)\cos\theta] }{ \sqrt{ \left[(\ell+1)^2-m^2\right] \left[(\ell+1)^2-s^2\right](\ell+s+2) } } \nonumber \\ & \left[ - \frac{m (\ell - s + 1) \cot\theta}{ [-m + (\ell + 1) \cos\theta] } - \frac{ (\ell + 1)^2 (s - 1) \cos\theta \cot\theta }{ (\ell - s + 1) [-m + (\ell + 1) \cos\theta] } \right. \nonumber \displaybreak \\ &\left. + \frac{ \left[s^2 (\ell + 1) - (\ell - m^2 + 1) (\ell + 1) - m^2 s\right] \csc\theta }{ (\ell - s + 1) [-m + (\ell + 1) \cos\theta] } \right. \nonumber \\ &\left. - \frac{ \ell (\ell + 1)^2 \sin\theta }{ (\ell - s + 1) [-m + (\ell + 1) \cos\theta] } \right] , \nonumber \\
%4
\alpha_{(s,\ell,m)\rightarrow(s+1,\ell-1,m)} &= \sqrt{\frac{2\ell-1}{2\ell+1}} \frac{ -(\ell + s) (m + \ell \cos\theta) }{ \sqrt{ (\ell^2 - m^2) (\ell^2 - s^2) (\ell + s) (\ell - s - 1) } } \left[ - \frac{ m (\ell + s) \cot\theta }{ (m + \ell\cos\theta) } \right. \nonumber \\ & \left. + \frac{ (\ell^2 - 2 m^2 - 2 \ell s) \csc\theta }{ 2(m + \ell\cos\theta) } - \frac{ \ell^2 \cos2\theta \csc\theta }{ 2(m + \ell\cos\theta) } \right].
\end{align}

Using the above relations, we obtain:
\begin{align} \label{raising_final}
\spSlmgq = ( \sbetalmgq \partial_\theta + \salphalmgq ) \sSlmgq + O(\g^2)
\end{align}
where
\begin{align}
\sbetalmgq =& \frac{-1}{\sqrt{(\ell-s)(\ell+s+1)}} - \frac{ m (\ell^2 + \ell + 2 s + 1) \g }{ \ell^2 (\ell+1)^2 \sqrt{(\ell - s) (\ell + s + 1)} } - \frac{ (2s+1)\cost \,\g}{ \ell (\ell+1) \sqrt{(\ell - s) (\ell + s + 1)} } \\
\salphalmgq =& 
\frac{m \csc\theta + s\cot\theta}{\sqrt{ (\ell-s)(\ell+s+1) }} 
+ \frac{ m \left[\ell (\ell + 1) + 3 \ell s (\ell + 1) + 2 s^2 + s\right] \cot\theta\, \g }{ \ell^2 (\ell+1)^2 \sqrt{(\ell - s) (\ell + s + 1)} } \nonumber \\ & 
+ \frac{\left[ \ell s(\ell+1)(2s+1) + m^2 (\ell^2+\ell+2s+1) \right]\csc\theta \, \g}{\ell^2(\ell+1)^2\sqrt{(\ell-s)(\ell+s+1)}} - \frac{\sin\theta\,\g}{\sqrt{(\ell-s)(\ell+s+1)}}.
%
%- \frac{ (2 s + 1) \cost \cot\theta\,\g }{ 2\ell (\ell+1) \sqrt{(\ell - s) (\ell + s + 1)}  } - \frac{ (2 \ell (\ell+1) + 2 s + 1)\sint\,\g }{ 2\ell (\ell+1) \sqrt{(\ell - s) (\ell + s + 1)}  } \nonumber \\ & + \frac{ (2 m^2 + \ell (\ell + 1) (2 m^2 + 1) + 4 s (\ell^2 + \ell + m^2) + 
%   4 \ell s^2 (\ell + 1)) \csc\theta\,\g }{ 2\ell^2 (\ell+1)^2 \sqrt{(\ell - s) (\ell + s + 1)} } .
\end{align}

Finally, using the relation,
\begin{align}
{\,\;S_{\ell,-m\!\!\!\!\!\!\!\!\!\!\!\!\!\!\!\!\!\!\!\!\!\!\!\!\!\!\!-s\,\,\,\,\,\,\,\,\,\,\,\,\,\,\,\,\,\,}^{-\gamma}} = (-1)^{m+s} \sSlmgq
\end{align}
we identify the lowering operator: 
\begin{align} \label{lowering_final}
{\,\;S_{\ell,m\!\!\!\!\!\!\!\!\!\!\!\!\!\!\!\!\!\!\!\!\!\!\!\!\!\!\!s-1\,\,\,\,\,\,\,\,\,\,\,\,\,\,\,\,\,\,}^{\gamma}} = -(\,\,\,\,{\,\;\beta_{\ell,-m\!\!\!\!\!\!\!\!\!\!\!\!\!\!\!\!\!\!\!\!\!\!\!\!\!\!\!-s\,\,\,\,\,\,\,\,\,\,\,\,\,\,\,\,\,\,\,\,}^{-\gamma}} \partial_\theta \,\,+\,\, {\,\;\alpha_{\ell,-m\!\!\!\!\!\!\!\!\!\!\!\!\!\!\!\!\!\!\!\!\!\!\!\!\!\!\!-s\,\,\,\,\,\,\,\,\,\,\,\,\,\,\,\,\,\,\,\,}^{-\gamma}}) \sSlmgq + O(\g^2).
\end{align}

%%%%%%%%%%%%%%%%%%%%%%%%%%%%%%%%%%%%%%%%%%%%%%%
%%%%%%%%%%%%%%%%%%%%%%%%%%%%%%%%%%%%%%%%%%%%%%%
%%%%%%%%%%%%   EXTRA SENTENCE FROM REVIEWER TWO    %%%%%%%%%%
%%%%%%%%%%%%%%%%%%%%%%%%%%%%%%%%%%%%%%%%%%%%%%%
%%%%%%%%%%%%%%%%%%%%%%%%%%%%%%%%%%%%%%%%%%%%%%%

\section{Conclusion}

Eqs (\ref{raising_final}) and (\ref{lowering_final}) represent the final result of our work -- equations for the linear-in-$\gamma$ spin-raising and lowering operators for spin-weighted spheroidal harmonics. We have obtained these, and eqs (62) and (63) for the required $\salphalmgq$ and $\sbetalmgq$, using methods described in Section \ref{sec2} and which can be readily extended to higher order in $\gamma$, if required.  Some special cases, corresponding to $s=\pm \ell$ and/or $m=\pm \ell$, are dealt with further in the Appendix.

\begin{acknowledgments}
This work was supported in part by the European Research Council under the European Union's Seventh Framework Programme (FP7/2007-2013)/ERC grant agreement no. 304978 to the University of Southampton, and NSF Grants PHY 1205906 and PHY 1314529 to the University of Florida.  Hospitality for BFW at the University of Southampton at an early stage in this work is gratefully acknowledged.  Support from the CNRS through the IAP, where part of this work was carried out, is also acknowledged, as well as support from the French state funds managed by the ANR within the Investissements d'Avenir programme under Grant No. ANR-11-IDEX-0004-02. 
\end{acknowledgments}

\appendix
\section{Building $\sTlmg$ for a special cases}
\label{sec5}

For cases such as $m=\ell$, $s=-\ell$ or $m=\ell=s$, one cannot simply use the general form,
\begin{align} \label{eqnA1}
\sTlmg = \sXlm  +  \g \, \sbllpmq \,\sXlpmq +  \g \, \sbllmmq \,\sXlmmq, 
\end{align}
with $\sbllmmq=0$ as when building $\sSlmgq$ for these cases. Instead, it then becomes imperative to carefully look at the analyticity of the Clebsch-Gordan coefficients in $\sbllmmq$, the coefficients of $\eth^s Q_{\ell-1}^m$, and at times even multiplying them with extra coefficients so that $\sTlmgq$ for such cases satisfy Eq. (\ref{sSlmg_eqn}). 

%When building spin-weighted spheroidal harmonics of type-II, the $\sTlmg$, for cases like $\ell=m$, $\ell=s$, etc, it is imperative to carefully look at the analyticity of the Clebsch-Gordan coefficients present in $\sbllmm$ and the coefficients of $\eth^s Q_\ell^{(m)}$. For example, though for $m>\ell$ we use $\sYlm=0$ (due to the Legendre type-I, $P_\ell^{(m)}=0$ for $m>\ell$) and $\sbllmm=0$, one cannot use the same for $\sXlm$ and the $\sbllmm$ multiplying it if $\sTlmg$ were to solve Eq. (\ref{sSlmg_eqn}). Moreover, one may need to multiply $\sbllmm \sXlmm$ with an extra coefficient for $\sTlmg$ to satisfy Eq. (\ref{sSlmg_eqn}).

In general cases, it is straight forward to use 
\begin{align}
{\,\;b_{\ell,\ell'\!\!\!\!\!\!\!\!\!\!\!\!\!\!\!s\,\,\,\,\,\,\,\,\,\,\,\,\,\,\,\,\,\,\,}^m} &= 2s \left( \frac{2\ell+1}{2\ell^\prime+1} \right)^{1/2} \frac{\left< \ell,m;1,0 | \ell^\prime,m\right>  \left< \ell,-s;1,0 | \ell^\prime,-s \right>  }{\ell(\ell+1) - \ell^\prime(\ell^\prime+1)}, \, \textrm{and}
\end{align}
\begin{align}\label{sXlm}
\sXlm(\theta) &= \sqrt{\frac{(2\ell+1)(\ell-m)!}{4\pi(\ell+m)!}} \,\, \eth_s\,\, Q_\ell^{m} (\cos\theta)
\end{align}
where $\left<|\right>$ are the Clebsch-Gordan coefficients which can be calculated using
\begin{align}
&\left< j_1,m_1 ; j_2,m_2 | j,m \right> = \sqrt{\frac{(j_1 + j_2 - j)! (j + j_1 - j_2)! (j + j_2 - j_1)! (2 j + 1)}{(j + j_1 + j_2 + 1)!}} \nonumber \\ &\times \sum_{k=-\infty}^\infty  \frac{(-1)^k \sqrt{(j_1 + m_1)! (j_1 - m_1)! (j_2 + m_2)! (j_2 - m_2)! (j + m)! (j - m)!}}{k! (j_1 + j_2 - j - k)! (j_1 - m_1 - k)! (j_2 + m_2 - k)! (j - j_2 + m_1 + k)! (j - j_1 - m_2 + k)!} .
\end{align} 

Now, we study the following special cases.

%%%%%%%%%%%%%%%%%%%%%%%%%%%%%%%%%%%%%%%%%%%%%%%%%%%%
%%%%%%%%%%%%%%%%%%%%%%%%%%%%%%%%%%%%%%%%%%%%%%%%%%%%
%%%%%%%%%%%%%%%%%%%%%%%%%%%%%%%%%%%%%%%%%%%%%%%%%%%%
%%%%%%%%%%%%%%%%%%%%%%%%%%             m = ell           and  | s | != ell 
%%%%%%%%%%%%%%%%%%%%%%%%%%%%%%%%%%%%%%%%%%%%%%%%%%%%
%%%%%%%%%%%%%%%%%%%%%%%%%%%%%%%%%%%%%%%%%%%%%%%%%%%%
%%%%%%%%%%%%%%%%%%%%%%%%%%%%%%%%%%%%%%%%%%%%%%%%%%%%
\subsection{$m = \ell$ and $|s| \ne \ell$}

In this case, there is a cancellation in the last term of the right-hand-side of Eq. (\ref{eqnA1}). The $\sbllmmq$ includes $\left< \ell,m;1,0 | \ell-1,m\right>$ which go to zero as 
\begin{align}
\frac{(2m-1)}{\sqrt{2(2m+1)}}\sqrt{(\ell-m)} + O(\ell-m)^{3/2}.
\end{align}
And $\sqrt{(\ell-1-m)!}$ in the numerator of the square-root of Eq. (\ref{sXlm}) blows up as $\frac{1}{\sqrt{(\ell-m)}}$ which cancels the one above.  Using these and after empirically finding the coefficient $c_\ell$, we have
\begin{align}
\sTlmgq = \sXlmq + \g \, \sbllpmq \,\,\sXlpmq + \g \, c_{\ell} \,\sbtildellmmq \,\,\sXtildelmmq + O(\g^2)
\end{align}
where
\begin{align} %\allowdisplaybreaks
c_\ell &= \frac{-2}{2\ell-1}, \nonumber \\
\sbtildellmmq &= 2s\left( \frac{2\ell+1}{2\ell-1} \right)^{1/2} \frac{\frac{2m-1}{\sqrt{2(2m+1)}}\left< \ell,-s;1,0 | \ell-1,-s\right>}{\ell(\ell+1)-\ell(\ell-1)}, \, \textrm{and}\nonumber  \\
\sXtildelmmq &= \sqrt{\frac{2\ell-1}{4\pi(\ell-1+m)!}} \, \eth^s Q_\ell^{m}.
\end{align}

%%%%%%%%%%%%%%%%%%%%%%%%%%%%%%%%%%%%%%%%%%%%%%%%%%%%
%%%%%%%%%%%%%%%%%%%%%%%%%%%%%%%%%%%%%%%%%%%%%%%%%%%%
%%%%%%%%%%%%%%%%%%%%%%%%%%%%%%%%%%%%%%%%%%%%%%%%%%%%
%%%%%%%%%%%%%%%%%%%%%%%%%%             m = -ell           and  | s | != ell
%%%%%%%%%%%%%%%%%%%%%%%%%%%%%%%%%%%%%%%%%%%%%%%%%%%%
%%%%%%%%%%%%%%%%%%%%%%%%%%%%%%%%%%%%%%%%%%%%%%%%%%%%
%%%%%%%%%%%%%%%%%%%%%%%%%%%%%%%%%%%%%%%%%%%%%%%%%%%%
\subsection{$m = -\ell$ and $|s| \ne \ell$}
The $\sbllmmq$ includes $\left< \ell,m;1,0 | \ell-1,m\right>$ which goes to zero as 
\begin{align}
-\sqrt{\frac{2}{1-2m}}\sqrt{(\ell+m)} + O(\ell+m)^{3/2}.
\end{align}
And $\sqrt{(\ell-1+m)!}$ in the denominator of the square-root of Eq. (\ref{sXlm}) goes to zero as $\sqrt{(\ell+m)}$. $\eth^s Q_{\ell}^{m}$ for $m=-\mu$ and $\ell = \mu-1$ (where $\mu > 0$) blows up to leading order as 
\begin{align}
(-1)^s  \frac{ \sqrt{(\mu+s-1)!(\mu-s-1)!} \, \tan^s(\theta/2) Q_{\mu-1}^{\mu}(\cos\theta) }{ (\mu-1)! \, ((2\mu-1)!)^{3/2} (\ell+m)^{3/2}}.
\end{align}
The coefficient $c_{\ell}$ is found empirically to be $(-1)^{\ell} \sqrt{(2\ell-1)!} \sqrt{(\ell+m)}$. Using these we get
\begin{align}
\sTlmgq = \sXlmq + \g \,  \sbllpmq \, \sXlpmq + \g \, c_\ell \, \sbtildellmmq \,\, \sXtildelmmq + O(\g^2)
\end{align}
where
\begin{align}
c_\ell &= (-1)^{\ell} \sqrt{(2\ell-1)!}, \\
\sbtildellmmq &= 2s\left( \frac{2\ell+1}{2\ell-1} \right)^{1/2}\frac{ -\sqrt{\frac{2}{1-2m}} \left< \ell,-s;1,0 | \ell-1,-s\right>}{\ell(\ell+1)-\ell(\ell-1)}, \, \textrm{and} \\
\sXtildelmmq &= (-1)^s \sqrt{\frac{(2\ell-1)(\ell-m-1)!}{4\pi}} \frac{ \sqrt{(s-m-1)!(-m-s-1)!} \, \tan^s(\theta/2) Q_{-m-1}^{-m}(\cos\theta) }{ (-m-1)! \, ((-2m-1)!)^{3/2} }
\end{align}
where appropriate powers of $\sqrt{\ell+m}$ in each of $c_\ell$, $\,_s\tilde{b}^m_{\ell,\ell-1}$ and $\,_s\tilde{X}_{\ell-1,m}$ %cancel each other to give a finite contribution. 
have been cancelled out to give a finite contribution.

%%%%%%%%%%%%%%%%%%%%%%%%%%%%%%%%%%%%%%%%%%%%%%%%%%%%
%%%%%%%%%%%%%%%%%%%%%%%%%%%%%%%%%%%%%%%%%%%%%%%%%%%%
%%%%%%%%%%%%%%%%%%%%%%%%%%%%%%%%%%%%%%%%%%%%%%%%%%%%
%%%%%%%%%%%%%%%%%%%%%%%%%%             | m | != ell            and  s = ell
%%%%%%%%%%%%%%%%%%%%%%%%%%%%%%%%%%%%%%%%%%%%%%%%%%%%
%%%%%%%%%%%%%%%%%%%%%%%%%%%%%%%%%%%%%%%%%%%%%%%%%%%%
%%%%%%%%%%%%%%%%%%%%%%%%%%%%%%%%%%%%%%%%%%%%%%%%%%%%
\subsection{$|m| \ne \ell$ and $s=\ell$}
The $\sbllmmq$ includes $\left< \ell,-s;1,0 | \ell-1,-s\right>$ which, in the limit $\ell\rightarrow s$, goes to zero as 
\begin{align}
-\sqrt{\frac{2}{1+2s}}\sqrt{(\ell-s)} + O(\ell-s)^{3/2},
\end{align}
and $\eth^s Q_{\ell-1}^m(\theta)$ blows up as %$O(\sqrt{\ell-s})$.
\begin{align}
\frac{(-1)^m}{2^s} \sqrt{\frac{(2s-2)!!}{(2s-1)!!}} \frac{(s+m-1)!}{(s-1)!}  \frac{\csc^{s+m}\left(\frac{\theta}{2}\right) \sec^{s-m} \left( \frac{\theta}{2}\right)}{\sqrt{\ell-s}}.
\end{align}
Factors of $\sqrt{\ell-s}$ in the above two equations cancel each other and $\sTlmgq$ is then given by 
\begin{align}
\sTlmgq = \sXlmq + \g \, \sbllpmq \, \sXlpmq + \g \, \sbtildellmmq \,\, \sXtildelmmq + O(\g^2)
\end{align}
where
\begin{align}
\sbtildellmmq &= 2s\left( \frac{2\ell+1}{2\ell-1} \right)^{1/2}\frac{ -\sqrt{\frac{2}{1+2s}} \left< \ell,m;1,0 | \ell-1,m\right>}{\ell(\ell+1)-\ell(\ell-1)}, \, \textrm{and} \\
\sXtildelmmq &= \sqrt{\frac{(2\ell-1)(\ell-1-m)!}{4\pi(\ell-1+m)!}} \frac{(-1)^m}{2^s} \sqrt{\frac{(2s-2)!!}{(2s-1)!!}} \frac{(s+m-1)!}{(s-1)!} \csc^{s+m}\left(\frac{\theta}{2}\right) \sec^{s-m} \left( \frac{\theta}{2}\right).
\end{align}

%%%%%%%%%%%%%%%%%%%%%%%%%%%%%%%%%%%%%%%%%%%%%%%%%%%%
%%%%%%%%%%%%%%%%%%%%%%%%%%%%%%%%%%%%%%%%%%%%%%%%%%%%
%%%%%%%%%%%%%%%%%%%%%%%%%%%%%%%%%%%%%%%%%%%%%%%%%%%%
%%%%%%%%%%%%%%%%%%%%%%%%%%             | m | != ell           and  s = -ell
%%%%%%%%%%%%%%%%%%%%%%%%%%%%%%%%%%%%%%%%%%%%%%%%%%%%
%%%%%%%%%%%%%%%%%%%%%%%%%%%%%%%%%%%%%%%%%%%%%%%%%%%%
%%%%%%%%%%%%%%%%%%%%%%%%%%%%%%%%%%%%%%%%%%%%%%%%%%%%
\subsection{$|m| \ne \ell$ and $s = -\ell$}
The $\sbllmmq$ includes $\left< \ell,-s;1,0 | \ell-1,-s\right>$ which, in the limit $\ell\rightarrow -s$, goes to zero as 
\begin{align}
\frac{-(2s+1)}{\sqrt{2}{\sqrt{1-2s}}} \sqrt{(\ell+s)} + O(\ell+s)^{3/2},
\end{align}
and $\eth^s Q_{\ell-1}^m(\theta)$ blows up as %$O(\sqrt{\ell+s})$ (doesn't have a general form for this case). 
\begin{align}
\frac{(-1)^s}{2^{|s|}} \sqrt{\frac{(2|s|-2)!!}{(2|s|-1)!!}} \frac{(|s|+m-1)!}{(|s|-1)!} \frac{ \csc^{|s|-m} \left( \frac{\theta}{2} \right) \sec^{|s|+m} \left( \frac{\theta}{2} \right)  }{\sqrt{\ell+s}}
\end{align}
The $\sqrt{\ell+s}$ in the above two equations cancel each other and $\sTlmgq$ is then given by 
\begin{align}
\sTlmgq = \sXlmq + \g \, \sbllpmq \, \sXlpmq + \g \, c_\ell \, \sbtildellmmq \sXtildelmmq + O(\g^2)
\end{align}
where
\begin{align} \allowdisplaybreaks
c_\ell &=  \frac{-2}{2\ell-1} \nonumber \\
\sbtildellmmq &= 2s\left( \frac{2\ell+1}{2\ell-1} \right)^{1/2}\frac{ \frac{-(2s+1)}{\sqrt{2}{\sqrt{1-2s}}} \left< \ell,m;1,0 | \ell-1,m\right>}{\ell(\ell+1)-\ell(\ell-1)}, \, \textrm{and}  \\
\sXtildelmmq &= \sqrt{\frac{(2\ell-1)(\ell-1-m)!}{4\pi(\ell-1+m)!}} \frac{(-1)^s}{2^{|s|}} \sqrt{\frac{(2|s|-2)!!}{(2|s|-1)!!}} \frac{(|s|+m-1)!}{(|s|-1)!}  \csc^{|s|-m} \left( \frac{\theta}{2} \right) \sec^{|s|+m} \left( \frac{\theta}{2} \right)  .
\end{align}

%%%%%%%%%%%%%%%%%%%%%%%%%%%%%%%%%%%%%%%%%%%%%%%%%%%%
%%%%%%%%%%%%%%%%%%%%%%%%%%%%%%%%%%%%%%%%%%%%%%%%%%%%
%%%%%%%%%%%%%%%%%%%%%%%%%%%%%%%%%%%%%%%%%%%%%%%%%%%%
%%%%%%%%%%%%%%%%%%%%%%%%%%             m = ell           and  s = ell
%%%%%%%%%%%%%%%%%%%%%%%%%%%%%%%%%%%%%%%%%%%%%%%%%%%%
%%%%%%%%%%%%%%%%%%%%%%%%%%%%%%%%%%%%%%%%%%%%%%%%%%%%
%%%%%%%%%%%%%%%%%%%%%%%%%%%%%%%%%%%%%%%%%%%%%%%%%%%%
\subsection{ $m=\ell$ and $s=\ell$ }
This case is a hybrid form of the $s=\ell$ and the $m=\ell$ cases. The two Clebsch-Gordan coefficients in $\sbllmmq$, and the denominator in the square-root of the coefficient of $\eth^s Q_{\ell-1}^m$ have the same form as the $s=\ell$ and $m=\ell$ cases. The coefficient $c_\ell$ is also same as the one in the $m=\ell$ case. $\eth^s Q_{\ell-1}^m$ blows up as 
\begin{align}
(-1)^s\frac{\sqrt{(2s-1)!}}{2\sqrt{\ell-s}}\csc^{2s}(\theta/2)
\end{align}
which cancels the Clebsch-Gordan coefficient containing $s$. We then have
\begin{align}
\sTlmg = \sXlmq + \g \, \sbllpmq \, \sXlpmq + \g \, c_\ell \, \sbtildellmmq\,\,\sXtildelmmq
\end{align}
where
\begin{align}
c_\ell &= \frac{-2}{2\ell-1}, \nonumber \\
\sbtildellmmq &= 2s\left( \frac{2\ell+1}{2\ell-1} \right)^{1/2} \frac{\frac{2m-1}{\sqrt{2(2m+1)}} \frac{-\sqrt{2}}{\sqrt{2s+1}}}{\ell(\ell+1)-\ell(\ell-1)}, \, \textrm{and}\nonumber \\
\sXtildelmmq &= \sqrt{\frac{2\ell-1}{4\pi(\ell-1+m)!}} \, (-1)^s\frac{\sqrt{(2s-1)!}}{2}\csc^{2s}(\theta/2).
\end{align}

%%%%%%%%%%%%%%%%%%%%%%%%%%%%%%%%%%%%%%%%%%%%%%%%%%%%
%%%%%%%%%%%%%%%%%%%%%%%%%%%%%%%%%%%%%%%%%%%%%%%%%%%%
%%%%%%%%%%%%%%%%%%%%%%%%%%%%%%%%%%%%%%%%%%%%%%%%%%%%
%%%%%%%%%%%%%%%%%%%%%%%%%%             m = ell           and  s = -ell
%%%%%%%%%%%%%%%%%%%%%%%%%%%%%%%%%%%%%%%%%%%%%%%%%%%%
%%%%%%%%%%%%%%%%%%%%%%%%%%%%%%%%%%%%%%%%%%%%%%%%%%%%
%%%%%%%%%%%%%%%%%%%%%%%%%%%%%%%%%%%%%%%%%%%%%%%%%%%%
\subsection{$m=\ell$ and $s=-\ell$}
This case is a hybrid form of the $s=-\ell$ and the $m=\ell$ cases. The two Clebsch-Gordan coefficients in $\sbllmmq$, and the denominator in the square-root of the coefficient of $\eth^s Q_{\ell-1}^m$ have the same form as the $s=-\ell$ and $m=\ell$ cases. $\eth^s Q_{\ell-1}^m$ blows up as 
\begin{align}
(-1)^s\frac{\sqrt{(-2s-1)!}}{2\sqrt{\ell+s}}\sec^{-2s}(\theta/2)
\end{align}
which cancels the Clebsch-Gordan coefficient containing $s$. We then have
\begin{align}
\sTlmgq = \sXlmq + \g \, \sbllpmq \, \sXlpmq + \g \, c_\ell \, \sbtildellmmq \,\, \sXtildelmmq
\end{align}
where
\begin{align}
c_\ell &= \left(\frac{-2}{2\ell-1}\right)^2, \nonumber \\
\sbtildellmmq &= 2s\left( \frac{2\ell+1}{2\ell-1} \right)^{1/2} \frac{\frac{2m-1}{\sqrt{2(2m+1)}} \frac{-(2s+1)}{\sqrt{2}\sqrt{1-2s}}}{\ell(\ell+1)-\ell(\ell-1)}, \, \textrm{and}\nonumber  \\
\sXtildelmmq &= \sqrt{\frac{2\ell-1}{4\pi(\ell-1+m)!}} \, (-1)^s\frac{\sqrt{(-2s-1)!}}{2}\cos^{2s}(\theta/2).
\end{align}

%%%%%%%%%%%%%%%%%%%%%%%%%%%%%%%%%%%%%%%%%%%%%%%%%%%%
%%%%%%%%%%%%%%%%%%%%%%%%%%%%%%%%%%%%%%%%%%%%%%%%%%%%
%%%%%%%%%%%%%%%%%%%%%%%%%%%%%%%%%%%%%%%%%%%%%%%%%%%%
%%%%%%%%%%%%%%%%%%%%%%%%%%             m = -ell           and  s = ell
%%%%%%%%%%%%%%%%%%%%%%%%%%%%%%%%%%%%%%%%%%%%%%%%%%%%
%%%%%%%%%%%%%%%%%%%%%%%%%%%%%%%%%%%%%%%%%%%%%%%%%%%%
%%%%%%%%%%%%%%%%%%%%%%%%%%%%%%%%%%%%%%%%%%%%%%%%%%%%
\subsection{$m=-\ell$ and $s=\ell$} 
This case is a hybrid of the $m=-\ell$ and the $s=\ell$ cases. We use $\epsilon$ to denote the small factor $(\ell+m)$ [or $(\ell-s)$].  The two Clebsch-Gordan coefficients in $\sbllmmq$ go to zero as $\epsilon^{1/2}$ each, and the $\sqrt{(\ell-1+m)!}$ in the expression of $\sXlmm$ goes to zero as $\epsilon^{1/2}$. $\eth^s Q_{\ell-1}^m$ blows up as $\epsilon^{-2}$ and we empirically find that the coefficient $\,_sc_{\ell}$ is $(-1)^s \sqrt{(2\ell-1)!} \epsilon^{1/2}$ canceling the extra $\epsilon^{-1/2}$ left in $\eth^s Q_{\ell-1}^m$. Here $\epsilon$ is the small factor $(\ell+m)$ [or $(\ell-s)$].  We then have 
\begin{align}
\sTlmgq = \sXlmq + \g \, \sbllpmq \, \sXlpmq + \g \, \,_sc_\ell \, \sbtildellmmq \,\, \sXtildelmmq
\end{align}
where
\begin{align}
\,_sc_\ell &= (-1)^s\sqrt{(2\ell-1)!}, \nonumber \\
\sbtildellmmq &= 2s\left( \frac{2\ell+1}{2\ell-1} \right)^{1/2} \frac{\frac{-\sqrt{2}}{\sqrt{1-2m}} \frac{-\sqrt{2}}{\sqrt{2s+1}}}{\ell(\ell+1)-\ell(\ell-1)}, \, \textrm{and}\nonumber \\
\sXtildelmmq &= \sqrt{\frac{(2\ell-1)(\ell-1-m)!}{4\pi}} \, \frac{1}{2(2s-1)!}\sec^{2s}(\theta/2).
\end{align}

%%%%%%%%%%%%%%%%%%%%%%%%%%%%%%%%%%%%%%%%%%%%%%%%%%%%
%%%%%%%%%%%%%%%%%%%%%%%%%%%%%%%%%%%%%%%%%%%%%%%%%%%%
%%%%%%%%%%%%%%%%%%%%%%%%%%%%%%%%%%%%%%%%%%%%%%%%%%%%
%%%%%%%%%%%%%%%%%%%%%%%%%%             m = -ell           and  s = -ell
%%%%%%%%%%%%%%%%%%%%%%%%%%%%%%%%%%%%%%%%%%%%%%%%%%%%
%%%%%%%%%%%%%%%%%%%%%%%%%%%%%%%%%%%%%%%%%%%%%%%%%%%%
%%%%%%%%%%%%%%%%%%%%%%%%%%%%%%%%%%%%%%%%%%%%%%%%%%%%
\subsection{$m=-\ell$ and $s=-\ell$} 
Identical cancellations take place as in the previous case. We have
\begin{align}
\sTlmgq = \sXlmq + \g \, \sbllpmq \,  \sXlpmq + \g \, \,_sc_\ell \, \sbtildellmmq \,\, \sXtildelmmq
\end{align}
where
\begin{align} \allowdisplaybreaks
\,_sc_\ell &= (-1)^{s+1}\frac{2\sqrt{(2\ell-1)!}}{(2\ell-1)}, \nonumber \\
\sbtildellmmq &= 2s\left( \frac{2\ell+1}{2\ell-1} \right)^{1/2} \frac{\frac{-\sqrt{2}}{\sqrt{1-2m}} \frac{-(1+2s)}{\sqrt{2}\sqrt{1-2s}}}{\ell(\ell+1)-\ell(\ell-1)}, \, \textrm{and}\nonumber \\
\sXtildelmmq &= \sqrt{\frac{(2\ell-1)(\ell-1-m)!}{4\pi}} \, \frac{1}{2(2|s|-1)!}\csc^{2|s|}(\theta/2).
\end{align}

\bibliography{sSlm}

\end{document}